\documentclass[10pt,twocolumn]{article}

\usepackage[utf8]{inputenc}
\usepackage[T1]{fontenc}
\usepackage{mathptmx}              % Times New Roman font
\usepackage{microtype}             % Gelişmiş tipografi
\usepackage{hyphenat}              % \nohyphenation komutu için

\usepackage{amsmath, amssymb, amsthm}

\usepackage{graphicx}
\usepackage{booktabs}
\usepackage{siunitx}
\usepackage{array}
\usepackage{ragged2e}
\usepackage{indentfirst}

\usepackage{hyperref}

\usepackage[a4paper, margin=1in]{geometry}

\usepackage{fancyhdr}
\usepackage{titlesec}
\titleformat{\section}
  {\centering\bfseries\fontsize{12}{14}\selectfont}
  {\thesection}{1em}{}
\titleformat{\subsection}
  {\centering\bfseries\fontsize{11}{13}\selectfont}
  {\thesubsection}{1em}{}
\titleformat{\subsubsection}
  {\centering\bfseries\fontsize{11}{13}\selectfont}
  {\thesubsubsection}{1em}{}

\usepackage{authblk} % Required for multiple authors with different affiliations

\title{Super-Resolution for Interferometric Imaging: Model Comparisons and Performance Analysis}

\author[1]{Hasan Berkay ABDIOGLU}
\author[1]{Rana GURSOY}
\author[1]{Yagmur ISIK}
\author[1]{Ibrahim Cem BALCI}
\author[2]{\\Taha UNAL}
\author[1]{Kerem BAYER}
\author[1]{Mustafa Ismail INAL} 
\author[4]{Nehir SERIN}
\author[1]{\\Muhammed Furkan KOSAR}
\author[3]{Gokhan Bora ESMER}
\author[1]{Huseyin UVET}

\affil[1]{Department of Mechatronics Engineering, Yildiz Technical University, Istanbul, Turkey}
\affil[2]{Department of Electronics and Communication Engineering, Yildiz Technical University, Istanbul, Turkey}
\affil[3]{Department of Electrical and Electronics Engineering, Marmara University, Istanbul, Turkey}
\affil[4]{Department of Department of Control and Automation Engineering, Yildiz Technical University, Istanbul, Turkey}

\begin{document}

\maketitle

\begin{abstract}
This study investigates the application of Super-Resolution techniques in holographic microscopy to enhance quantitative phase imaging. An off-axis Mach-Zehnder interferometric setup was employed to capture interferograms. The study evaluates two Super-Resolution models, RCAN and Real-ESRGAN, for their effectiveness in reconstructing high-resolution interferograms from a microparticle-based dataset. The models were assessed using two primary approaches: image-based analysis for structural detail enhancement and morphological evaluation for maintaining sample integrity and phase map accuracy. The results demonstrate that RCAN achieves superior numerical precision, making it ideal for applications requiring highly accurate phase map reconstruction, while Real-ESRGAN enhances visual quality and structural coherence, making it suitable for visualization-focused applications. This study highlights the potential of Super-Resolution models in overcoming diffraction-imposed resolution limitations in holographic microscopy, opening the way for improved imaging techniques in biomedical diagnostics, materials science, and other high-precision fields.
\end{abstract}

\vspace{1ex} % biraz boşluk ekler

\section{Introduction}
Holographic microscopy is a label-free imaging technique that allows for high-resolution quantitative phase imaging. The system functions by capturing holograms via optical interference, producing an interferogram that encodes spatially varying phase information as interference fringes \cite{Goodman1996}. This interferogram reflects variations in the optical path length across the sample, with each pixel containing intensity values resulting from optical interference.  However, inaccuracies in phase reconstruction, whether due to noise, limited resolution, or suboptimal data processing, can significantly affect the reliability of the extracted information \cite{Park2018}.

Despite its advantages, holographic microscopy is inherently constrained by diffraction limits, as dictated by the Rayleigh criteria. According to this, the minimum resolvable feature size is given by the equation:

\begin{equation}
r_{\text{min}} = \frac{1.22\,\lambda}{D}\\[5pt]
\end{equation}

where \(\lambda\) represents the illumination wavelength and \(D\) denotes the aperture diameter \cite{Goodman1996}. This diffraction-imposed limitation restricts the finest details that can be resolved and directly affects the accuracy of phase extraction. If resolution constraints are not addressed, errors in phase mapping may propagate through subsequent analyzes, leading to distortions in height measurements, displacement calculations, and overall quantitative phase imaging \cite{Li2023}.

Given that resolution is a critical factor in microscopy, significant research efforts have been directed towards surpassing diffraction-induced limitations. Super-Resolution models have emerged as a promising computational approach to enhance image resolution beyond the diffraction limit \cite{Wang2019}. These techniques employ signal processing and machine learning methodologies to reconstruct high-resolution images from lower-resolution datasets. Super-Resolution methods not only enhance image quality, but also provide a cost-effective means to extract better-quality data without requiring modifications to the physical optical setup. By computationally overcoming physical resolution limitations, Super-Resolution techniques can significantly extend the capabilities of holographic microscopy, making it more accessible for high-precision applications \cite{Rivenson2017}.

The effectiveness of Super-Resolution methods in interferometric imaging depends on the quality of the initial data. Experimental inconsistencies, including focus misalignment, insufficient illumination, and surface contaminants, can introduce artifacts that compromise phase accuracy. A properly trained Super-Resolution model, utilizing high-quality reference datasets, may reduce these effects by diminishing noise, enhancing contrast, and improving resolution and phase map reconstruction \cite{Liu2019}. 

In this study, we developed a deep learning–based Super-Resolution model using a holographic microscopy dataset acquired from polyacrylamide and agarose-based microparticles with known dimensions. This approach standardizes the data while preserving the integrity of the phase maps. Our findings demonstrate that deep learning–driven Super-Resolution techniques can overcome inherent diffraction limits and enhancing lateral resolution of holographic microscopy.

\section{Methodology}

\subsection{Data Collection System}

\begin{figure}[ht]
    \centering
    \vspace{-0.3cm}
    \includegraphics[width=\linewidth]{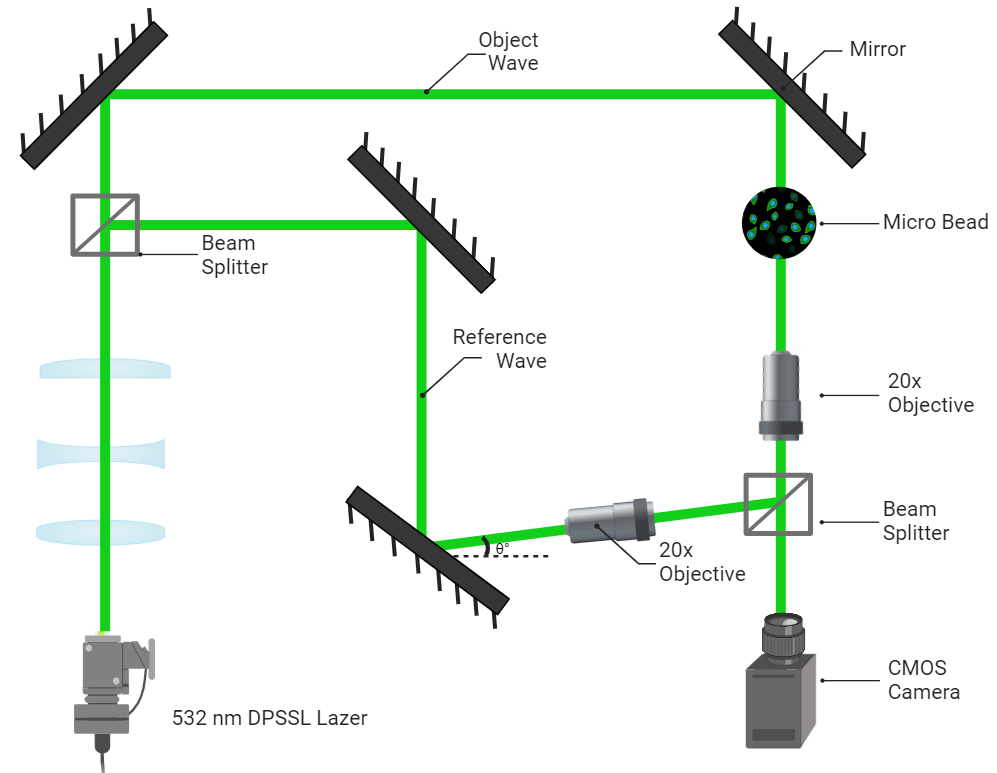}
    \vspace{-0.8cm}
    \caption{\centering Schematic Representation of the Mach-Zehnder Interferometer Setup}
    \label{fig:set_up}
\end{figure}

In this study an off-axis Mach-Zehnder interferometric setup was used for optical imaging. A coherent laser beam was split into two paths: the object beam, which interacted with the microparticle, and the reference beam, which remained unaltered (Fig.\ref{fig:set_up}). These beams were then recombined at a beam splitter, generating an interferogram that encoded phase variations in the form of interference fringe patterns, also referred to as phase lines. The off-axis configuration introduced a controlled angular offset, spatially separating interference orders in the Fourier domain, thereby enabling single-shot phase map retrieval via Fourier transform techniques.

The captured holograms consisted of single-channel intensity images, with each pixel representing the local light intensity. Numerical reconstruction involved isolating the region of interest, applying a two-dimensional Fast Fourier Transform (FFT) to extract phase information, and subsequently performing an inverse FFT to reconstruct the complex amplitude of the object wave \cite{poon2014introduction}.

The initial phase measurements were confined within a limited range, resulting in abrupt discontinuities in phase values. To achieve a continuous phase distribution, phase unwrapping techniques were employed, utilizing an arctangent-based method to adjust the phase at each pixel. This process ensured accurate accounting of phase changes occurring in multiples of \(2\pi\). Additionally, distortions caused by background noise and low-frequency variations were mitigated through polynomial surface fitting, thereby improving the accuracy of the reconstructed phase map \cite{Vannoni2007}. 

The phase map represents the morphological surface of the observed microparticles in terms of phase \cite{Marquet2005}. After the unwrapping process, these phase data are transformed into an optical path difference (OPD) map, which corresponds to height measurements derived from phase information. This transformation is performed using the established relationship:\cite{Vannoni2007}:
\begin{equation}
\text{OPD} = \frac{\lambda}{2\pi} \phi
\end{equation}
where \( \lambda \) is the laser wavelength. This transformation facilitated a quantitative assessment of the sample's morphological structure.

The conversion of angular phase data into metric measurements enabled a more meaningful quantitative comparison of morphological features. Throughout the study, this method was consistently applied to enhance result interpretability and ensure a comprehensive and accurate assessment of the sample's structural characteristics.

\subsection{Data Preprocessing and Degradation Handling}

Interferometric imaging is inherently affected by multiple sources of degradation, including sensor noise, optical aberrations, and environmental fluctuations, all of which introduce distortions that can impact phase retrieval accuracy. A controlled degradation procedure was implemented to maintain the dataset's suitability for computational modeling while keeping the authenticity of the phase map, balancing realism and methodological consistency.

\begin{figure}[ht]
    \centering
    \vspace{-0.3cm}
    \includegraphics[width=\linewidth]{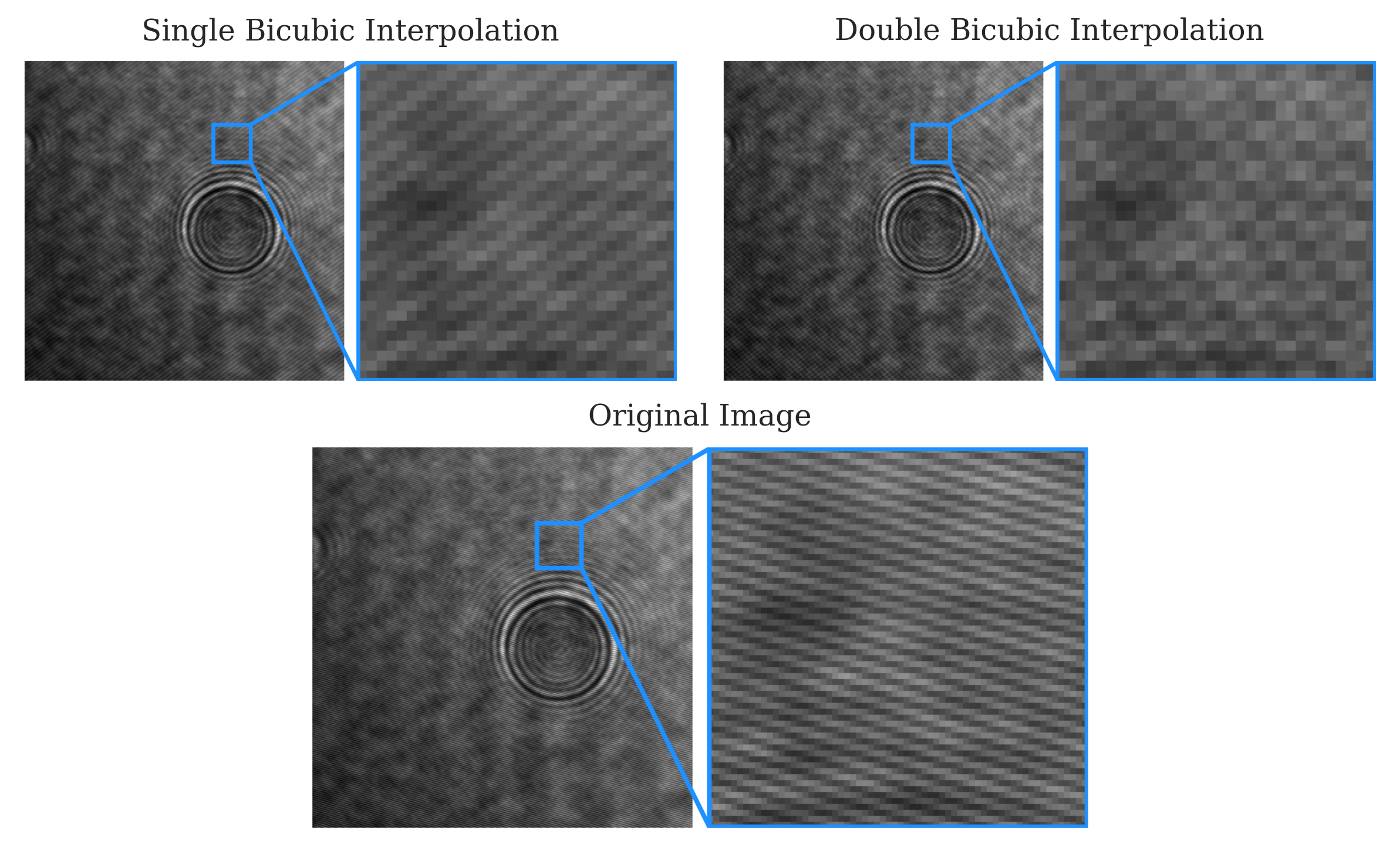}
    \vspace{-0.8cm}
    \caption{\centering Outputs of Different Downscaling Methodologies }
    \label{fig:downscale}
\end{figure}

A key aspect of this process was the adoption of a two-stage interpolation-based downscaling strategy, instead of a single-step approach (Fig.\,\ref{fig:downscale}). Bicubic interpolation was utilized in both methodologies. A direct bicubic interpolation with a downscaling factor of 2 was utilized in the single-step method. The two-stage approach employed two consecutive bicubic interpolations, each with a downscaling factor of \(\sqrt{2}\). Experimental analysis revealed that while single-step interpolation introduced undesirable directional distortions in phase lines, it also preserved their visibility. In contrast, the two-stage method effectively reduced these distortions while simultaneously decreasing phase line visibility. This reduction was controlled to ensure that the model could still learn effectively without misinterpretation. Ultimately, the two-stage approach achieved a balance, minimizing distortion while retaining sufficient detail for accurate model training.

To further refine the dataset and enhance the complexity of the learning process, Gaussian noise was systematically introduced as an additional degradation step, with its amplitude adjusted between 1\% and 5\%. In contrast to the inherent noise generated by the imaging system, this deliberate degradation was applied to eliminate residual phase-line artifacts, thereby ensuring that the model was trained on a more diverse and unbiased dataset.

By integrating two-stage downscaling strategy and noise addition, the dataset was optimized to more accurately reflect real-world interferometric imaging conditions while preserving the integrity of phase map matrices \cite{JiRealSR}.

\subsection{Architecture of Super-Resolution Models}
To evaluate the applicability of Super-Resolution models for interferogram image reconstruction, two models with distinct approaches were selected: RCAN and Real-ESRGAN. Although there are numerous Super-Resolution models, this study focuses on training one CNN-based model and one GAN-based model, each offering unique advantages in processing interferograms.

CNN-based models excel at capturing spatial relationships and local features. Their relatively fast training process facilitates the development of deeper networks without significantly increasing computational costs \cite{Lecun2015,Lecun1998}. Among CNN-based Super-Resolution models, RCAN was chosen due to its deep yet efficient architecture, which preserves fine details while maintaining computational feasibility \cite{Zhang2018}.

In contrast, GAN-based models can generate highly realistic and visually convincing results by leveraging an adversarial learning framework. These models have demonstrated remarkable performance in restoring missing or distorted details in interferograms, thereby enhancing structural integrity and improving overall image quality \cite{Geyer2023}. Among GAN-based Super-Resolution models, Real-ESRGAN was chosen due to its optimizations for synthetic low resolution data. Given that the dataset used in this study consists of synthetic low resolution images, Real-ESRGAN was selected as the GAN-based model \cite{Wang2021}.

\subsubsection{RCAN: Residual Channel Attention Network}
RCAN (Residual Channel Attention Network) is a deep learning–based architecture designed to address Super-Resolution tasks by preserving fine details in images. It employs residual connections and channel attention mechanisms to enhance the extraction of high-frequency features, thereby improving both learning efficiency and reconstruction quality \cite{Zhang2018}. Our method excludes the channel attention component due to the use of single-channel images.

A key component of RCAN is its residual-in-residual (RIR) structure, which facilitates the construction of extremely deep networks while mitigating the vanishing gradient problem. The model integrates long skip connections at the group level and short skip connections within residual blocks, ensuring that low-frequency information bypasses computationally intensive layers. This architectural design enables the network to focus on high-frequency details essential for accurate and visually convincing image reconstruction \cite{Zhang2018}.

Despite its depth, RCAN remains computationally efficient due to its effective residual block design, which promotes stable gradient propagation. Recent studies have demonstrated that, when trained with modern optimization techniques, RCAN performs on par with or even surpasses contemporary CNN-based Super-Resolution models \cite{Lin2022}. On the basis of these advantages, RCAN was selected as the primary CNN-based model for evaluating the suitability of Super-Resolution techniques in holographic microscopy applications.

\subsubsection{Loss Functions for RCAN Training}
To optimize the performance of RCAN, a combination of L1 loss and Multi-Scale Structural Similarity Index Measure (MS-SSIM) loss was used to balance pixel-wise precision with visual quality Both loss functions were assigned equal weights (1:1) during training to maintain equilibrium between structural consistency and numerical accuracy.

L1 Loss, also known as the mean absolute error (MAE), quantifies the average absolute difference between predicted and actual pixel values. It exhibits reduced sensitivity to outliers than L2 loss, making it particularly effective for preserving edges and fine textures. This characteristic is especially crucial in Super-Resolution tasks, where maintaining image sharpness and local details is essential \cite{Arezoomand2021}.

MS-SSIM Loss provides a similarity measure by evaluating image brightness, contrast, and structural coherence on multiple scales. Incorporating this loss function results in more realistic and structurally coherent reconstructions, aligning closely with human visual perception. Additionally, it mitigates oversmoothing, a common issue in Super-Resolution models that rely exclusively on pixel-wise loss functions\cite{Arezoomand2021}.

By integrating L1 and MS-SSIM losses, RCAN was optimized to generate high-fidelity, super-resolved images. This dual-loss strategy effectively balances numerical accuracy with visual consistency, addressing challenges such as excessive smoothing and the loss of fine details.

\subsubsection{Real-ESRGAN: Enhanced Super-Resolution Generative Adversarial Network}
Real-ESRGAN extends the Enhanced SuperResolution Generative Adversarial Networks (ESRGAN) model to more effectively handle real-world distortions. While both ESRGAN and Real-ESRGAN are trained on synthetically downscaled images, ESRGAN struggles with real-world degradations such as blurring, noise, and compression artifacts. In contrast, Real-ESRGAN incorporates higher-order degradation modeling, allowing it to better simulate these real-world distortions and achieve improved image restoration quality when applied to real-world data \cite{Wang2021}.

Real-ESRGAN comprises a generator and a discriminator. The generator reconstructs fine details, textures, and sharpness from low-quality inputs, while the discriminator assesses how closely the output resembles a real high-resolution image and provides feedback to refine the generator performance. Through adversarial training, the generator progressively enhances its outputs, producing increasingly realistic and visually convincing reconstructions \cite{Wang2021}.

Architecturally, the generator builds upon ESRGAN’s Residual-in-Residual Dense Blocks (RRDB), facilitating efficient information flow and mitigating gradient loss to improve stability and performance. Furthermore, it integrates a high-order distortion model to more effectively address real-world degradations and preserve fine textures \cite{Wang2021}.

\subsubsection{Loss Functions for Real-ESRGAN Training}
For the training of Real-ESRGAN, the loss functions used in the RCAN, L1 loss and the MS-SSIM loss were also used with equal weights (1:1). In addition, an adversarial loss, referred to as GAN loss, was incorporated to enhance visual quality. GAN loss originates from the adversarial training process, in which the generator and the discriminator compete to improve the realism of generated images \cite{Goodfellow2014}.

Although GAN loss is not the primary optimization objective, it plays a crucial role in reducing artifacts and enhancing image sharpness. In models trained with L2 loss, GAN loss has been shown to mitigate the tendency to produce overly smooth images. While L2 loss was not utilized in this study, GAN loss was assigned a relatively low weight to maintain training stability while still refining fine image details \cite{BulatSuperRes}.

This multi-loss strategy ensures that Real-ESRGAN generates high-fidelity reconstructions that preserve numerical accuracy while achieving visually realistic and convincing results.

\subsubsection{Training Optimization Strategies for Improved Performance}
To reduce training time and enhance model performance, modern optimization techniques were employed throughout the training process. One key approach was increasing the batch size and patch size to fully utilize the high VRAM capacity of modern GPUs. Using the parallel processing capabilities of GPUs, this strategy allowed more efficient training while maintaining computational feasibility \cite{Goyal2017}.

In this study, as proposed, the learning rate was scaled proportionally to the increased batch size. This adjustment ensured stable convergence and prevented potential issues such as vanishing or exploding gradients \cite{You2017}.

For optimization, the Adan optimizer was selected due to its superior convergence speed and its ability to dynamically adjust the learning rate for larger batch and patch sizes. Adan has demonstrated improved stability compared to conventional optimizers, as it is less sensitive to hyperparameter selection, thus promoting more consistent learning outcomes \cite{Xie2022}.

By applying these optimization strategies, the training process was accelerated while maintaining high reconstruction quality. This ensured that the model effectively enhanced interferograms, preserving structural integrity and fine details with improved efficiency.

\subsection{Evaluation of Super-Resolution Image Quality}

Interferometric measurements inherently capture only intensity information; however, Fourier analysis enables the extraction of a phase map, which reveals the morphological characteristics of the sample. To ensure a comprehensive assessment of image quality, an effective evaluation framework must integrate both intensity (visible) and frequency (phase) components. Accordingly, the proposed framework employs single-channel, pixel-level measurements for reconstructed image evaluation. In this context, the performance metrics are categorized into two primary groups: structural integrity metrics and error metrics \cite{Arabboev2024}.

\subsubsection{Structural Integrity Metrics}

Structural integrity metrics assess the visual and structural fidelity of reconstructed images by capturing luminance, contrast, and structural information.

The Structural Similarity Index Measure (SSIM) evaluates local image regions based on luminance, contrast, and structure, providing a visually relevant measure of structural coherence. This metric quantifies the degree to which the reconstructed image preserves structural details relative to the original:
\begin{equation}
SSIM(x,y) = [l(x,y)]^\alpha \cdot [c(x,y)]^\beta \cdot [s(x,y)]^\gamma,
\end{equation}
where \( l(x,y) \), \( c(x,y) \), and \( s(x,y) \) correspond to luminance, contrast, and structure comparisons, respectively, and \( \alpha \), \( \beta \), and \( \gamma \) are parameters that control their relative importance \cite{Arabboev2024}.

The Peak Signal-to-Noise Ratio (PSNR), measured in decibels (dB), evaluates the relationship between the maximum achievable signal and the reconstruction error. This metric delivers an intuitive numerical assessment of image quality by measuring the extent of signal degradation:
\begin{equation}
PSNR = 20 \cdot \log_{10}(MAX_I) - 10 \cdot \log_{10}(MSE),
\end{equation}
where \( MAX_I \) denotes the maximum possible pixel value and \( MSE \) represents the mean squared error. \cite{Arabboev2024}.

\subsubsection{Error Metrics}

Error metrics quantify the numerical differences between the reconstructed and ground truth images at the pixel level.

Mean Squared Error (MSE) quantifies the average squared difference between the ground truth and the reconstructed image. The Root Mean Squared Error (RMSE) is simply the square root of MSE:
\begin{equation}
RMSE = \sqrt{MSE} = \sqrt{\frac{1}{N} \sum_{i=1}^{N} \left( I_{GT,i} - I_{SR,i} \right)^2},
\end{equation}
where \( N \) is the total number of pixels, \( I_{GT,i} \) represents the intensity at pixel \( i \) in the ground truth image, and \( I_{SR,i} \) denotes the corresponding intensity in the reconstructed image \cite{Chai2014}.

Mean Absolute Error (MAE) calculates the average absolute difference between corresponding pixel intensities of the ground truth and reconstructed images. This metric provides a reliable assessment of the accuracy of reconstruction due to its uniform treatment of errors \cite{Chai2014}:
\begin{equation}
MAE = \frac{1}{N} \sum_{i=1}^{N} \left| I_{GT,i} - I_{SR,i} \right|.
\end{equation}

\section{Results}
Below are the visual results comparing the outputs of the RCAN and Real-ESRGAN models. These images provide a clear illustration of each model's reconstruction capabilities.

\begin{figure}[ht]
    \centering
    \vspace{-0.3cm}
    \includegraphics[width=\linewidth]{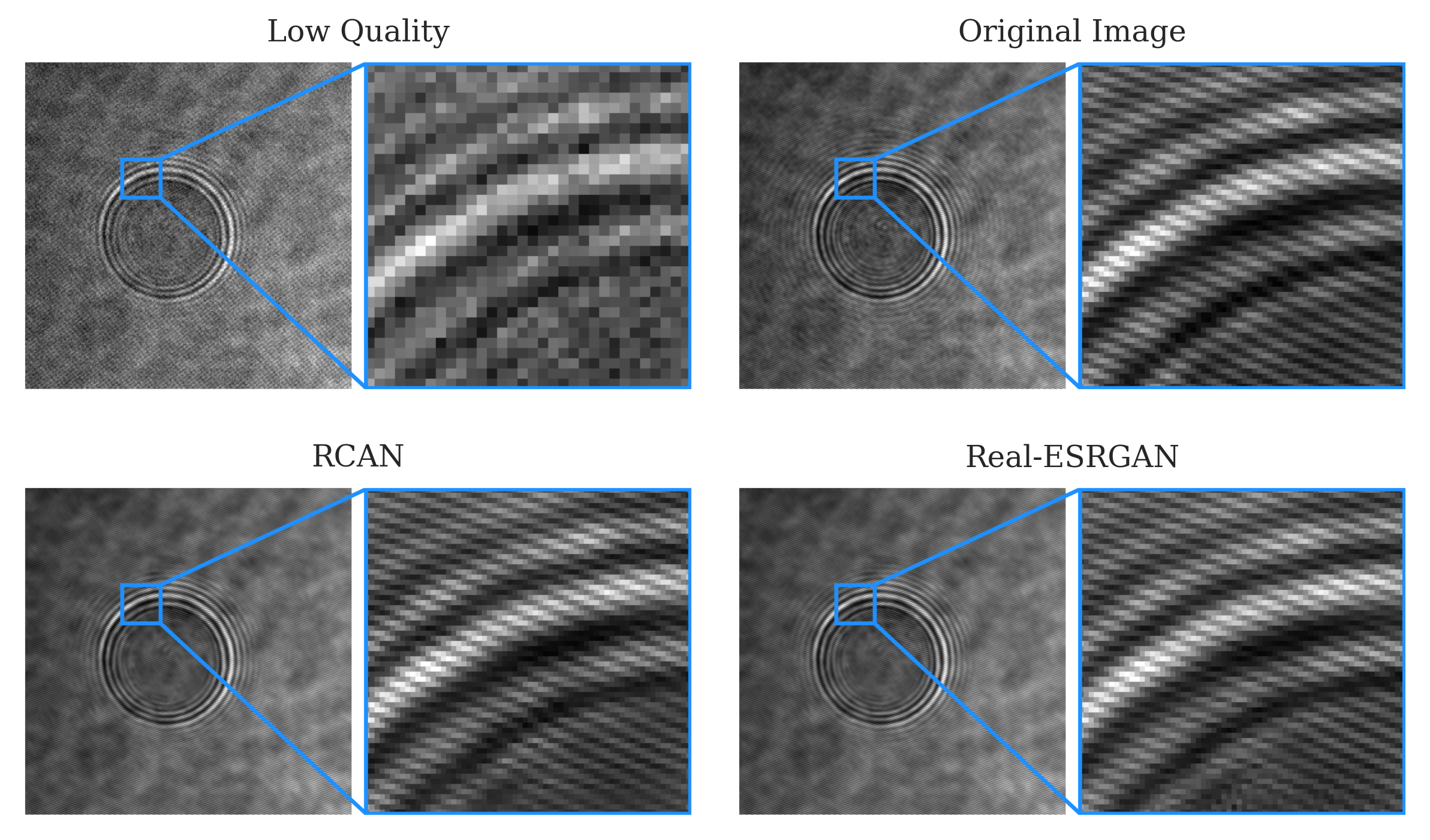}
    \vspace{-0.8cm}
    \caption{\centering Enlarged Visualization of the  Super-Resolution Results }
    \label{results}
\end{figure}

Also, interpreting quantitative metrics is crucial for evaluating the performance of trained models in image Super-Resolution. The selected framework components establish an objective foundation for comparing reconstruction accuracy and visual quality. Tables \ref{tab:my_table} and \ref{tab:model_performance} present the evaluation results for RCAN and Real-ESRGAN, providing insights into their effectiveness in generating high-quality images while accounting for morphological aspects of the reconstructions.

\subsection{Evaluation on 2D Image Data}

As shown in Table \ref{tab:my_table}, Real-ESRGAN consistently exhibited superior performance compared to RCAN across multiple evaluation metrics. It achieves lower reconstruction errors, as evidenced by the reduced MSE (0.0007416 vs. 0.0009034), RMSE (0.02616 vs. 0.02814), and MAE (0.02013 vs. 0.02246). Moreover, Real-ESRGAN attained a higher PSNR (31.95 dB vs. 31.50 dB), indicating improved fidelity, and a marginally enhanced SSIM (0.9633 vs. 0.9619), demonstrating superior structural preservation. These results suggest that the adversarial training incorporated in Real-ESRGAN contributes to more accurate reconstructions, particularly in preserving fine details and textures.

\begin{table}[ht]
    \caption{Performance Comparison of RCAN and Real-ESRGAN on the Test Dataset}
    \centering
    \vspace{+0.2cm}
    \begin{tabular}{>{\RaggedRight\arraybackslash}p{2.5cm} *{2}{c}}
        \toprule
        \textbf{Metric} & \textbf{RCAN} & \textbf{Real-ESRGAN} \\
        \midrule
        MSE & \num{0.0009034} & \num{0.0007416} \\
        RMSE & \num{0.02814} & \num{0.02616} \\
        MAE & \num{0.02246} & \num{0.02013} \\
        PSNR & \num{31.50} & \num{31.95} \\
        SSIM & \num{0.9619} & \num{0.9633} \\
        \bottomrule
    \end{tabular}
    \label{tab:my_table}
\end{table}

\begin{figure}[ht]
    \centering
    \vspace{-0.3cm}
    \includegraphics[width=\linewidth]{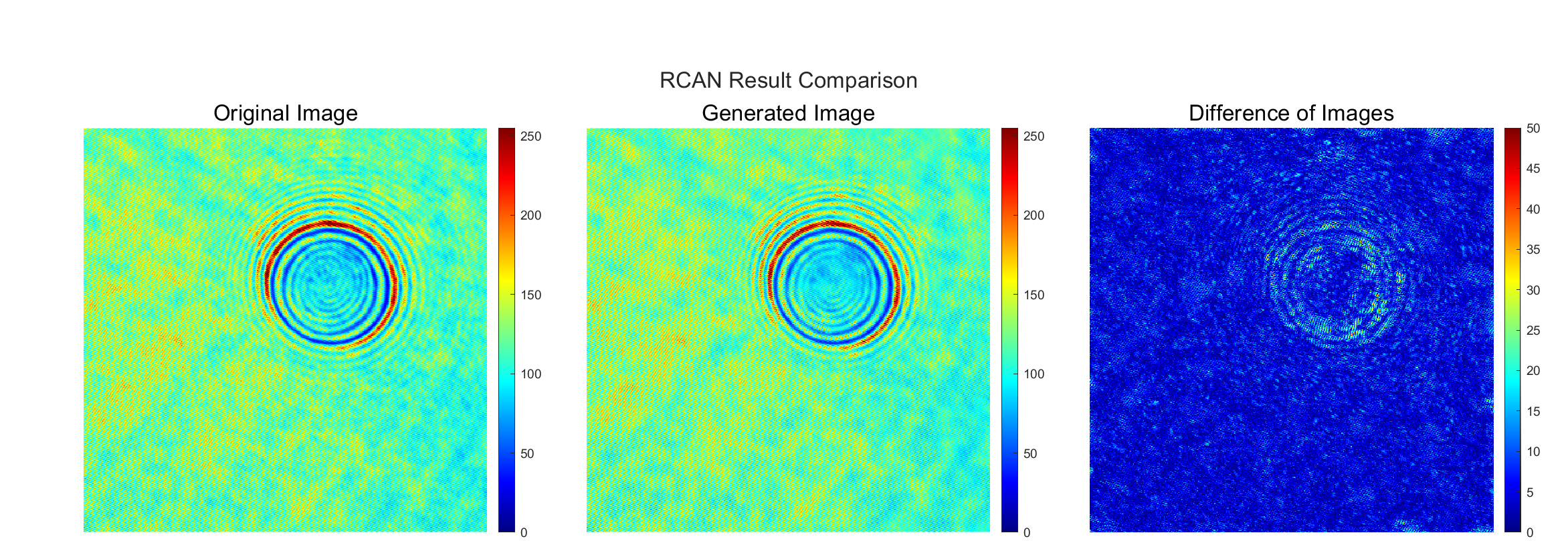}
    \includegraphics[width=\linewidth]{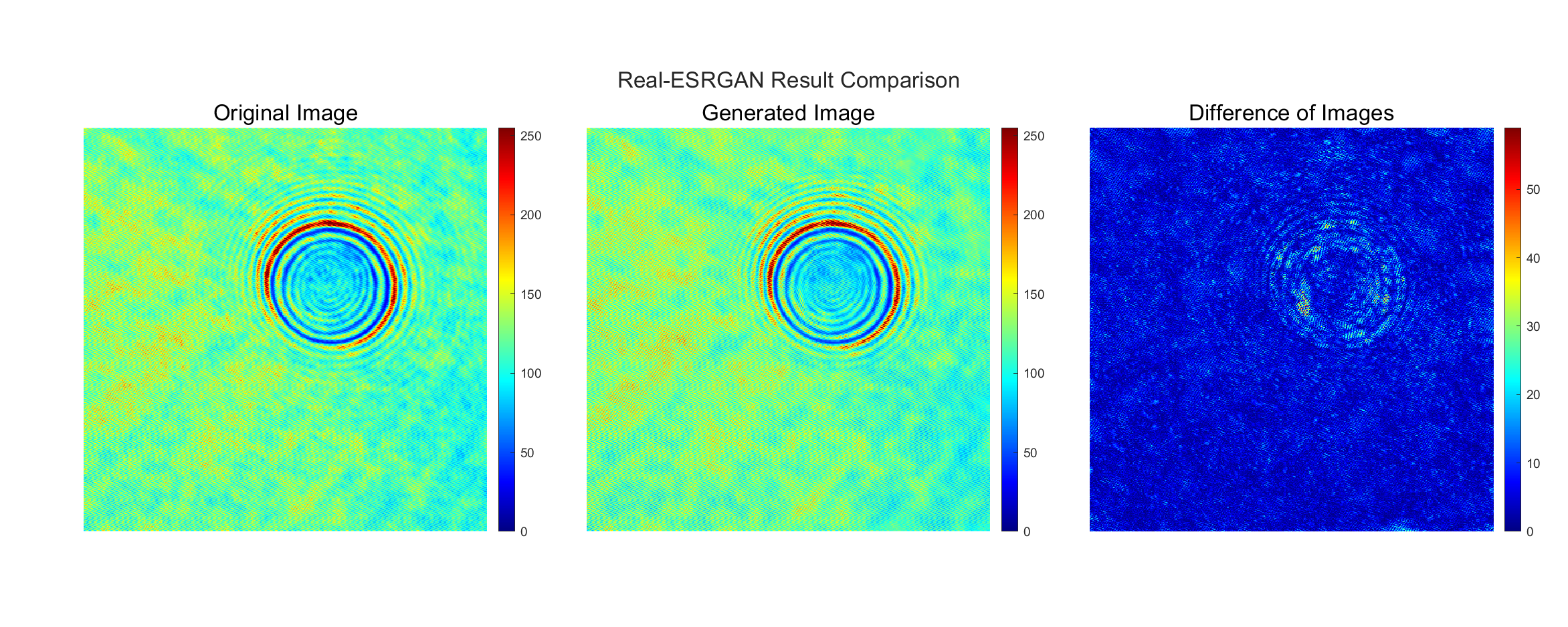}
    \vspace{-0.8cm}
    \caption{\centering Comparison of Super-Resolution and Original Images}
    \label{result_comp}
\end{figure}

\subsection{Evaluation of the Phase Data}

\begin{table}[ht]
\caption{RCAN and Real-ESRGAN Performance Comparison with Phase Data on the Test Dataset}
\centering
\vspace{+0.2cm}
\begin{tabular}{>{\RaggedRight\arraybackslash}p{2.5cm} *{2}{c}}
\toprule
\textbf{Metric} & \textbf{RCAN} & \textbf{Real-ESRGAN} \\
\midrule
MSE & $\num{0.5820}$ & $\num{0.7040}$ \\
RMSE & $\num{0.6665}$ & $\num{0.7069}$ \\
MAE & $\num{0.5203}$ & $\num{0.5263}$ \\
PSNR & $\num{5.029}$ & $\num{5.491}$ \\
SSIM & $\num{0.6249}$ & $\num{0.6419}$ \\
\bottomrule
\end{tabular}
\label{tab:model_performance}
\end{table}

Table \ref{tab:model_performance} presents a comparative analysis of RCAN and Real-ESRGAN on phase results of hologram images, revealing the distinct strengths of each model. Real-ESRGAN achieves superior structural integrity, as evidenced by higher SSIM (0.6419 vs. 0.6249) and PSNR (5.491 vs. 5.029), indicating improved visual quality and consistency. Conversely, RCAN demonstrates more precise error minimization, with lower MSE (0.5820 vs. 0.7040), RMSE (0.6665 vs. 0.7069), and MAE (0.5203 vs. 0.5263), suggesting a higher level of numerical accuracy in phase-map reconstruction.

\begin{figure}[ht]
    \centering
    \vspace{-0.3cm}
    \includegraphics[width=\linewidth]{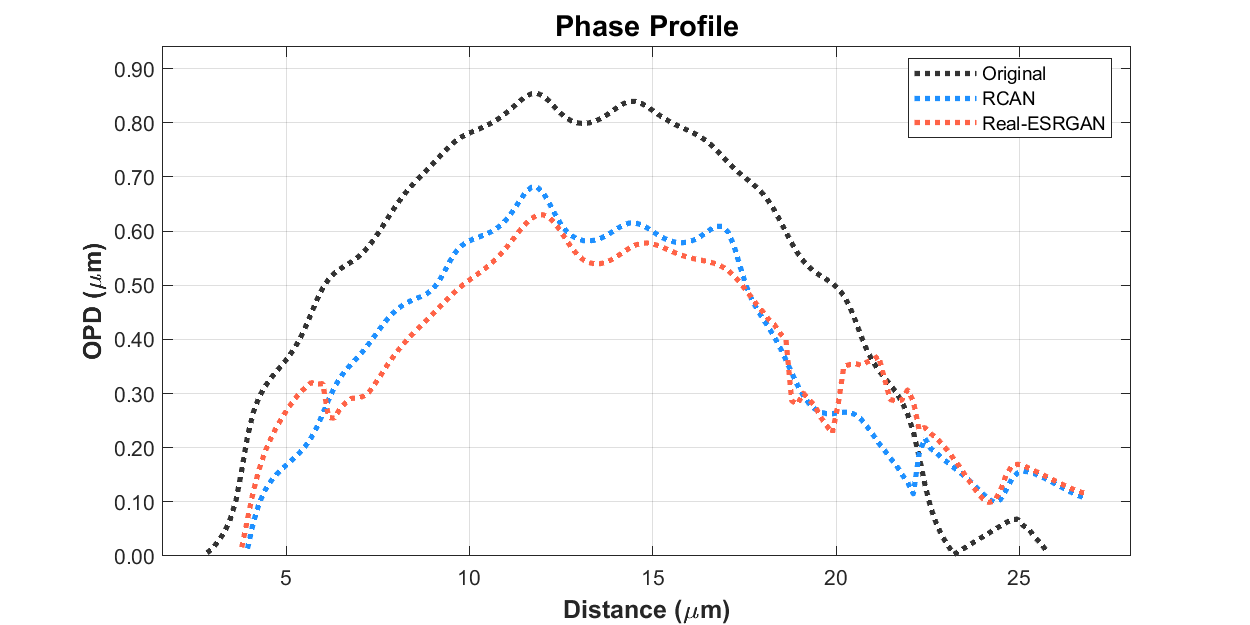}
    \vspace{-0.8cm}
    \caption{\centering A representative phase profile depicting the optical path difference (OPD) transformation applied to both the original and generated datasets. }
    \label{fig:graph}
\end{figure}

A notable difference between the phase image evaluation and previous image-based assessments is the substantial decline in PSNR and the increase in error metrics. This suggests that reconstructing phase images introduces additional challenges that are not present in hologram  images. However, despite this degradation, SSIM remained relatively intact, indicating that overall structural integrity was largely preserved.  Fig.\ref{fig:graph} illustrates that the reconstructed outputs exhibit a normalization effect, potentially due to the smoothing tendency in the Super-Resolution process.

The observed disparities in image-based and phase evaluations are due to the absence of phase-aware or morphology-based loss functions during training. These methods do not explicitly account for phase coherence because they maximize image quality simply using pixel-wise criteria. Integrating a phase-aware loss function can improve reconstruction fidelity by guaranteeing that interferograms maintain phase integrity.

Overall, these findings highlight the trade-off between numerical accuracy and visual quality. RCAN excels in minimizing reconstruction errors, making it preferable for applications that require precise quantitative analysis. In contrast, Real-ESRGAN produces more visually coherent results, making it better suited for applications in which visual quality and structural fidelity are prioritized.

\subsection{Model Selection Based on Application Requirements}

An important observation in the Real-ESRGAN generated images is the gradual reduction in surface imperfections as the number of iterations increases. This effect suggests that Real-ESRGAN, owing to its adversarial training approach, modifies reconstructed data to prioritize visual aesthetics over strict adherence to the original input. By enhancing visual quality and smoothing artifacts, the model effectively generates images that appear more natural but may introduce subtle modifications to the underlying structure. In contrast, RCAN focuses on preserving the fidelity of the original data because its architecture is designed to minimize numerical errors and maintain the integrity of fine details without significantly altering the input.

The choice between these models is based on the specific requirements of the intended application. If the objective is to enhance visual clarity, suppress noise, and produce visually pleasing images, Real-ESRGAN is a more suitable option, particularly when trained on a dataset tailored to the application. However, in cases where even minor alterations in the interferogram can lead to inaccurate analyses, RCAN is the preferred model because of its strict adherence to the input data and its emphasis on numerical accuracy.

Ultimately, selecting the appropriate model depends on balancing the visual enhancement and data fidelity. Applications focused on human perception, such as visualization tasks or artistic rendering, may benefit from the capabilities of Real-ESRGAN. Conversely, scientific and engineering applications requiring precise quantitative measurements require RCAN to preserve intricate details. This highlights the necessity of aligning the model’s characteristics with the specific needs of the application to ensure optimal performance and reliability of the reconstructed data.

\section{Conclusion}
The models developed in this study have demonstrated strong performance in enhancing interferometric imaging, particularly in improving phase map reconstruction accuracy and structural coherence. Real-ESRGAN consistently exhibited superior visual quality, achieving higher SSIM and PSNR values, making it preferable for visualization-driven applications. Meanwhile, RCAN provided higher numerical precision, with lower reconstruction errors, making it more suitable for applications requiring accurate quantitative analysis.

However, challenges were encountered in reconstructing phase data, as both models exhibited performance degradation compared to hologram image based assessments. This highlights the trade-offs between numerical accuracy and visual quality, with RCAN excelling in minimizing reconstruction errors and Real-ESRGAN producing more visually coherent results. The choice between these models depends on the specific application requirements, whether prioritizing structural precision or visual fidelity. These findings emphasize the importance of aligning Super-Resolution model selection with the intended use case to ensure optimal performance in interferometric imaging.

% ============================================================================
%                           Bibliografi Ayarları
% ============================================================================
\bibliographystyle{IEEEtran}
\bibliography{main}

% ============================================================================
\end{document}